\documentstyle[preprint,aps]{revtex}

\begin{document}

\draft
\title{		Formal Proof of an Exact Scale Invariance 
			in the Nambu-Jona-Lasinio Model}
\author{			Keiichi Akama}
\address{	Department of Physics, Saitama Medical College,
                     Kawakado, Moroyama, Saitama, 350-04, Japan}
\date{\today}
\maketitle

\begin{abstract}
We show that the renormalization group $\beta $ functions 
	in the Nambu-Jona-Lasinio model identically vanish in all order
	due to the compositeness condition.
Accordingly the effective coupling constants are entirely fixed
	and do not run with the renormalization scale.
\end{abstract}


Quantum fluctuations in some models give rise to composite fields
	by supplying them the kinetic terms \cite{NJL,QC}. 
The quantum composites well describe various physical phenomena
	and are widely applied to 
	the models of hadrons,
	the composite gauge bosons and
	composite Higgs scalar \cite{ATTCA}, 
	collective modes in the nuclear and condensed matter physics, etc..
They can be formulated as the special case of some renormalization theory
	with the compositeness condition (CC) \cite{CC},
	which says that $Z=0$, 
	where $Z$ is the wave-function renormalization constant 
	of the to-be-composite field.
For example, with CC, the Yukawa model for elementary fermions and bosons 
	reduces to the Nambu-Jona-Lasinio (NJL) type model \cite{NJL} 
	with elementary fermions and quantum composite bosons. 
Here we consider the renormalization group (RG) properties 
	of CC in the NJL model \cite{Wilson}. 
For definiteness, we consider the RG of the 
	't Hooft type throughout this paper.

The RG analyses with CC in the NJL model
	have had great impact in phenomenological models \cite{top}.
The theoretical aspects of the interplay between RG and CC
	have been also of continual interest of the people \cite{RG}.
Most of them, however, considered the limit of the infinite momentum cutoff, 
	which necessitates some additional uncertain assumptions,
	such as ladder approximation, non-perturbative fixed point, etc.,
	because perturbatively the NJL model is trivial at the limit.
Here we take the momentum cutoff 
	as a large but finite physical parameter, 
	and rely on  none of theese assumptions.
Then we prove the somewhat surprising but important theorem:
{\it In the NJL model, i.e. in the Yukawa model with CC, 
	the RG beta functions identically vanish 
	due to the CC itself,
	and consequently the coupling constants are scale-invariant}.
In the previous paper, we demonstrated it in the RG flow
	at the leading order in $1/N$ 
	where $N$ is number of the matter fermion species \cite{beta}.
The derivation suggests that it holds also in higher order.
In this paper, we present a simple argument 
	which formally proves the scale invariance {\it in all order}.
This theorem is important because it appears to contradict with 
	the widely spread use of CC and RG in phenomenology \cite{top}.
More importantly, however, 
	such a high symmetry is realized in a rather awkward cutoff theory. 
It is a new type of scale invariance.
Furthermore it is an all-order exact result 
	derived from the simple and general assumptions.

We consider the NJL model in its simplest form given by the Lagrangian 
\begin{eqnarray} 
	{\cal L}_{\rm N}=\overline \Psi i\!\not\!\partial \Psi 
	+f|\overline \Psi _{\rm L}\Psi _{\rm R}|^2,
	\label{LNb}
\end{eqnarray} 
where $\Psi=(\Psi _{1}, \Psi _{2}, \cdots , \Psi _{N})$ 
	is a bare color $N$-plet fermion, 
	$f$ is a bare coupling constant, 
	the subscripts ``L" and ``R" indicate chiralities.
Since the model is not renormalizable,
	we need to introduce some regularization scheme with a finite cutoff.
We adopt the dimensional regularization where we consider everything 
	in $d(=4-2\epsilon )$ dimensional spacetime 
	with small but non-vanishing $\epsilon $.
The parameter $\epsilon $ roughly corresponds to $1/\log\Lambda $ 
	with the momentum cutoff $\Lambda $.
The system is equivalent to that described by the 
	Lagrangian \cite{KK}
\begin{eqnarray} 
	{\cal L}'_{\rm N}=\overline \Psi i\!\not\!\partial \Psi 
	+(\overline \Psi _{\rm L}\Phi\Psi _{\rm R}+{\rm h.c.})
	-{1\over f}|\Phi |^2,
	\label{L'Nb}
\end{eqnarray} 
where $\Phi $ is an auxiliary field.

Now we compare this with the renormalizable Yukawa model 
	for the elementary fermion $\psi _0$ and the elementary boson $\phi _0$
	with the following Lagrangian
\begin{eqnarray} 
	{\cal L}_{\rm Y}=\overline \psi _0i\!\not\!\partial \psi _0
	+g_0(\overline \psi _{\rm 0L}\phi _0\psi _{\rm 0R}+{\rm h.c.})
	+|\partial _\mu \phi _0|^2
	-m_0^2|\phi _0|^2
	-\lambda _0|\phi _0|^4		\label{LY}
\end{eqnarray} 
where $m_0$ is the bare mass of $\phi _0$, 
	and $g_0$ and $\lambda _0$ are bare coupling constants. 
To absorb the divergences of the quantum loop diagrams, 
	we renormalize the fields, the mass, and the coupling constants as
\begin{eqnarray} 
&&	\psi _0= \sqrt {Z_\psi } \psi ,\ \ \   
	\phi _0= \sqrt {Z_\phi } \phi ,\ \ \   
	Z_\phi m_0^2= Z_m m^2, 
\\&&
	Z_\psi \sqrt {Z_\phi }g_0= {Z_g} g \mu ^\epsilon ,\ \ \ \ \  
	Z_\phi ^2\lambda _0= Z_\lambda  \lambda  \mu ^{2\epsilon },\ \ \ \ \  
\label{ZZ}
\end{eqnarray} 
where $\psi $, $\phi $, $m$, $g$, and $\lambda $ are the renormalized 
	fields, mass, and coupling constants, respectively, 
	$Z_\psi $, $Z_\phi $, $Z_m$, $Z_g$ and $Z_\lambda $ are
	the renormalization constants,
	and $\mu $ is a mass scale parameter 
	to make $g$ and $\lambda $ dimensionless. 
Then the Lagrangian ${\cal L}_{\rm Y}$ becomes
\begin{eqnarray} 
&&	{\cal L}_{\rm Y}=Z_\psi \overline \psi i\!\not\!\partial \psi 
	+Z_gg\mu ^\epsilon (\overline \psi _{\rm L}\phi \psi _{\rm R}+{\rm h.c.})
\cr &&\ \ \ \ \ \ \ \ \ \ \  
	+Z_\phi |\partial _\mu \phi |^2
	-Z_m m^2|\phi |^2
	-Z_\lambda \lambda \mu ^{2\epsilon }|\phi |^4.
		\label{LYR}
\end{eqnarray} 
As the renormalization condition, 
	we adopt the minimal subtraction scheme, 
	where, as the divergent part
	to be absorbed into in the renormalization constants, 
	we retain all the negative power terms in the Laurent 
	series in $\epsilon $ of the divergent (sub)diagrams.
Then the parameter $\mu $ is interpreted as the renormalization scale.
Since the coupling constants are dimensionless,
	the renormalization constants depend on $\mu $ 
	only through $g$ and $\lambda $, but do not explicitly depend on $\mu $.

Now we can see that the Lagrangian (\ref{LYR}) of the Yukawa model
	coincides with the Lagrangian (\ref{L'Nb}) of the NJL model, if 
\begin{eqnarray} 
	Z_\phi =Z_\lambda =0, \ \ 
	Z_\psi  \not= 0, \ \ 
	Z_g \not= 0, \ \ 
	Z_m \not= 0. \ \ 
	\label{CC}
\end{eqnarray} 
The condition (\ref{CC}) is the so-called ``compositeness condition" (CC) \cite{CC}
	which imposes relations among the coupling constants $g$ and $\lambda $, 
	the mass $m$, and the cutoff parameter $\epsilon $ in the Yukawa model 
	so that it reduces to the NJL model.
In both of the expansion in the coupling constants and that in $1/N$,
	the perturbative calculations show that 
	$g\rightarrow0$ and  $\lambda\rightarrow0$ 
	as $\epsilon\rightarrow0$ at each order,
	and the theory becomes trivial free theory.
Therefore we fix the cutoff $\Lambda=\mu e^{1/\epsilon}$ at some finite value.
We can read off from (\ref{L'Nb}) and (\ref{LYR}) 
	that the fields and parameters of the NJL and the Yukawa models
	should be connected by the relations
\begin{eqnarray} 
&&	\Psi=\sqrt{Z_\psi}\psi,\ \
	\Phi={Z_g g\mu^\epsilon \over Z_\psi}\phi,\ \
	f= {\ Z_g^2 g^2\mu ^{2\epsilon}
	/ Z_\psi^{2} Z_m m^{2}}.  
		\label{Cf}
\end{eqnarray} 
The last of (\ref{Cf}) is so-called ``gap equation" of the NJL model.
In terms of the bare parameters, 
	the CC (\ref{CC}) corresponds to the limit
\begin{eqnarray} 
	g_0\rightarrow \infty , \ \ \ \ \ 
	\lambda _0/g_0^4\rightarrow 0.\label{limbare}
\end{eqnarray} 
These behaviors may look singular at first sight,
	but they are of no harm 
	because they are unobservable bare quantities.

Thus the NJL model is equivalent to 
	the cutoff Yukawa model (i.e. the Yukawa model with a finite cutoff)
	under the CC (\ref{CC}).
Then the RG of the former 
	coincides with that of the latter under the CC (\ref{CC}).
Let us consider the latter (the cutoff Yukawa model)
	with special cares on the finite cutoff.
In our case, it amounts to fix $\epsilon =(4-d)/2$ at some non-vanishing value.
The beta functions and the anomalous dimensions are defined as 
\begin{eqnarray} &&
	\beta _g^{(\epsilon )}(g,\lambda )=\mu {\partial g\over \partial \mu }\ , \ \ \ \ 
	\beta _\lambda ^{(\epsilon )}(g,\lambda )=\mu {\partial \lambda \over \partial \mu }\ ,\label{betadef}\\
&&
	\gamma _\phi ^{(\epsilon )}(g,\lambda )={1\over 2}\mu {\partial \ln Z_\phi \over \partial \mu }\ , \ \ \ \ 
	\gamma _\psi ^{(\epsilon )}(g,\lambda )={1\over 2}\mu {\partial \ln Z_\psi \over \partial \mu }\ ,
\end{eqnarray} 
where the differentiation $\partial /\partial \mu $ performed with
	$g_0$, $\lambda _0$, and $\epsilon $ fixed.
Operating $\mu (\partial /\partial \mu )$ to the equations in (\ref{ZZ}) we obtain
\begin{eqnarray} 
	\left[ \beta _g^{(\epsilon )}{\partial \over \partial g}+\beta _\lambda ^{(\epsilon )}{\partial \over \partial \lambda }+\epsilon \right] gJ=0, \ \ \ \ 
	\left[ \beta _g^{(\epsilon )}{\partial \over \partial g}+\beta _\lambda ^{(\epsilon )}{\partial \over \partial \lambda }+2\epsilon \right] \lambda K=0,
\end{eqnarray} 
where $J=Z_g/(Z_\psi \sqrt {Z_\phi })$ and $K=Z_\lambda /Z_\phi ^2$.
Comparing the residues of the poles at $\epsilon =0$, we obtain
\begin{eqnarray} 
	\beta _g^{(\epsilon )}=-\epsilon g+g{\cal D}J_1, \ \ \ \  
	\beta _\lambda ^{(\epsilon )}=-2\epsilon \lambda +\lambda {\cal D}K_1, \ \label{beta}
\end{eqnarray} 
where ${\cal D}=g(\partial /\partial g)+2\lambda (\partial /\partial \lambda )$, and 
	$J_1$ and $K_1$ are the residues of the simple poles 
	of $J$ and $K$, respectively.
On the other hand the anomalous dimensions are given by
\begin{eqnarray} 
	\gamma _\phi ^{(\epsilon )}=-{1\over 2}{\cal D}Z_{\phi 1}, \ \ \ \ 
	\gamma _\psi ^{(\epsilon )}=-{1\over 2}{\cal D}Z_{\psi 1}, \label{gamma}
\end{eqnarray} 
where $Z_{\phi 1}$ and $Z_{\psi 1}$ are the residues of the simple poles 
	of $Z_{\phi }$ and $Z_{\psi }$, respectively.
We can read off from (\ref{beta}) and (\ref{gamma})
	that $\beta ^{(\epsilon )}$'s depend on the cutoff only through
	the first terms $-\epsilon g$ and $-2\epsilon \lambda $ of the expressions,
	while $\gamma ^{(\epsilon )}$'s are independent of $\epsilon $.
We should be careful not to neglect the cutoff dependence of $\beta ^{(\epsilon )}$'s.

For an illustration, we start with the leading-order approximation in $1/N$.
Explicit calculations give
\begin{eqnarray} 
	Z_\phi =1-{Ng^2\over 16\pi ^2\epsilon }\ ,\ \ \ 
	Z_\lambda =1-{Ng^4\over 16\pi ^2\epsilon \lambda }\ ,\ \ \ 
	Z_g=Z_\psi =1.\ \ \ \label{ZZ1}
\end{eqnarray} 
Using (\ref{beta}) and (\ref{gamma}), we get
\begin{eqnarray} &&
	\beta _g^{(\epsilon )}=-\epsilon g+{Ng^3\over 16\pi ^2}\ ,\ \ \ 
	\beta _\lambda ^{(\epsilon )}=-2\epsilon \lambda +{N(4g^2\lambda -2g^4)\over 16\pi ^2}\ ,\ \ \ \label{beta1}
\\&&
	\gamma _\phi ^{(\epsilon )}={Ng^2\over 16\pi ^2}\ ,\ \ \ 
	\gamma _\psi ^{(\epsilon )}=0.\ \ \ 
\end{eqnarray} 
The CC $Z_\phi =Z_\lambda =0$ in (\ref{CC}) with (\ref{ZZ1}) 
	is solved to give
\begin{eqnarray} 
	g^2=\lambda ={16\pi ^2\epsilon \over N}\ .\label{CCsol1}
\end{eqnarray} 
Then we substitute (\ref{CCsol1}) into (\ref{beta1}) 
	to get $\beta _g^{(\epsilon )}=\beta _\lambda ^{(\epsilon )}=0$. 
Thus the statement that the beta functions in the NJL model vanish
	is proved at this order.
The coupling constants $g$ and $\lambda $ do not run with the scale $\mu $.
In the previous paper, we demonstrated the scale invariance of the NJL model
	in the renormalization group flow of the general case \cite{beta}.

If we consider coupling-constant expansion or loop expansion
	instead of the $1/N$ expansion,
	the $\beta ^{(\epsilon )}$'s appear to fail to vanish.
It is, however, because the former expansions 
	(the coupling-constant and the loop expansions)
	are inconsistent in the NJL model as follows.
At the leading order in these expansions,
	the CC takes the same form as (\ref{CCsol1}).
The leading-order contributions to $\beta ^{(\epsilon )}$ involve 
	the diagrams with boson lines.
A one-fermion-loop insertion into a boson line give rise to 
	a extra factor of $O(g^2N/\epsilon )$, 
	which is order unity according to (\ref{CCsol1}).
Then the infinitely many higher-order diagrams 
	with one-fermion-loop insertions 
	have the same order of magnitude.
Therefore the coupling-constant and the loop expansions fail,
The suitable expansion is that in $1/N$.

Thus the NJL model is at a fixed point in the RG flow of the Yukawa model
	at this order.
The coupling constants in the NJL model are scale-invariant,
	and do not run with the scale parameter.
We can trace back the reason of scale invariance
	to the fact that beta functions vanish 
	due to the compositeness condition.
It is further traced back to the fact that the scale invariance of
	the relation  (\ref{ZZ}) under the compositeness condition (\ref{CC}).
Thus we expect that the scale invariance holds not only in 
	the leading order in $1/N$, but also in all order.

Now we show that, under the compositeness condition, 
the beta functions vanish in all order.
The dependence of $g$ and $\lambda  $ on $\varepsilon  $ and $\mu  $ should be 
	determined by the RG equations (\ref{betadef}) 
	with the beta functions in (\ref{beta}).
They are derived from eq. (\ref{ZZ}), 
	and it is  eq. (\ref{ZZ}) that originally determines
	the $\varepsilon $- and $\mu $-dependence of $g$ and $\lambda  $.
We rewrite (\ref{ZZ}) into the form
\begin{eqnarray} 
    F\equiv 	Z_\phi  -{Z_g^2 g^2
\over 
                Z_\psi  ^2}\   {\mu  ^{2\varepsilon  }
\over  
                              g_0^2   }=0\ ,\ \ \ \ 
    G\equiv 	Z_\lambda  -{Z_g^4 g^4
\over 
              Z_\psi  ^4\lambda      }\ {\lambda _0\mu  ^{2\varepsilon  }
\over 
                                   g_0^4         }=0
\ .\ \ \ \ \label{ZZG}
\end{eqnarray} 
Note that $Z$'s are functions of $g$, $\lambda $, and $\varepsilon $, and consequently 
	$F$ and $G$ are functions of $g$, $\lambda $, $\varepsilon $, $\mu $, $g_0$, and $\lambda _0$. 
A mathematical theorem says that, 
	if $\partial(F,G)/\partial(g,\lambda )\not= 0$ in an appropriate region,
	the equation (\ref{ZZG}) has a unique solution for $g$ and $\lambda $ 
	as functions of $\varepsilon $, $\mu $, $g_0$, and $\lambda _0$,  
	and 
\begin{eqnarray} 
\pmatrix{{\partial g \over \partial \mu }\cr {\partial \lambda \over \partial \mu }}
 =-\pmatrix{{\partial F\over \partial g}&{\partial F\over \partial \lambda }\cr 
 {\partial G\over \partial g}&{\partial G\over \partial \lambda }}^{-1}
 \pmatrix{{\partial F\over \partial \mu }\cr {\partial G\over \partial \mu }} .   \label{MMM}
\end{eqnarray}
It is sufficient for us to consider the case 
	where the CC (\ref{CC}) has a nontrivial solution.
Therefore $\partial(Z_\phi ,Z_\lambda )/\partial(g,\lambda )\not=0$, 
	$g\not=0$, and $\lambda\not=0$.
Then, by (\ref{ZZG}), the first matrix in r.h.s.\ of (\ref{MMM}) is finite
	in the NJL limit (\ref{limbare}).
On the other hand, in the same limit, we have 
\begin{eqnarray} 
    {\partial F\over \partial \mu }=-{Z_g^2 g^2\over  Z_\psi  ^2}\ {2\varepsilon \mu  ^{2\varepsilon -1}\over  g_0^2}\rightarrow 0,\ \ \ \ 
    {\partial G\over \partial \mu }=-{Z_g^4 g^4\over  Z_\psi  ^4\lambda  }\ {2\varepsilon \lambda  _0\mu  ^{2\varepsilon -1}\over  g_0^4}\rightarrow 0,
\ \ \ \ \label{dFG}
\end{eqnarray} 
Therefore, from (\ref{MMM}), $\partial g/\partial \mu \rightarrow 0$ and $\partial \lambda /\partial \mu \rightarrow 0$,
	and hence $\beta  _g^{(\varepsilon  )}\rightarrow 0$ and $\beta  _\lambda ^{(\varepsilon  )}\rightarrow 0$ in the compositeness limit.
Thus, {\it the compositeness condition
implies scale invariance of the coupling constants in all order}.

What is newly proved here is that 
	the NJL model is always just on the fixed point 
	of the cutoff Yukawa model in the all order. 
The often claimed statement 
	``there is `a' nontrivial infrared stable fixed point in the NJL model" 
	\cite{Wilson} is inappropriate 
	because no other possibility is allowed than the fixed point. 
The scale invariant (Yukawa) coupling constant is not ``a" solution, 
	but the only solution in the NJL model. 
It is fixed even in the ultraviolet region. 
We can demonstrate it in lower orders in $1/N$ 
	by using the compositeness condition 
	which characterizes the NJL model in the cutoff Yukawa model \cite{beta}. 
Here it is proved in the all order.

So far we have illustrated the theorem in terms of the simplest NJL model.
It is, however, obvious that the theorem can be extended to more general cases
	including e.g.\ a vectorial composite \cite{compG}, 
	a boson-boson or a boson-fermion composite.
It would be also applicable to the composite gravity \cite{preg} 
	and the brane induced gravity \cite{BIG}, 
	by using renormalizable $R^2$ gravity, if we ignore some drawbacks.
It is surprising that the non-renormalizable NJL model with
	a finite cutoff 	shares the property of vanishing $\beta $ function
	with highly symmetric models 
	like e.g.\ the $N=4$ super-Yang-Mills theory.
The CC of the NJL-type model
	is so strong as to impose such a high symmetry.

\end{document}